# Broadband microwave time–frequency analysis via stabilized period-one oscillation and recirculating frequency shifting with shared fiber loop

Xianxin Zhang, Chi Jiang, Taixia Shi, Yang Chen*

Shanghai Key Laboratory of Multidimensional Information Processing, School of Information and Electronic Engineering, East China Normal University, Shanghai, 200241, China
*ychen@ce.ecnu.edu.cn

**ABSTRACT**
To address the need for time–frequency analysis (TFA) of broadband microwave signals—and to overcome the reliance of existing schemes on large-bandwidth swept microwave sources—we propose a broadband microwave signal TFA approach based on stabilized period-one (P1) oscillation and recirculating frequency shifting (RFS). By employing a feedback loop, a stable swept optical signal is generated from the stabilized P1 oscillation of a semiconductor laser, and its sweep bandwidth is further extended by an RFS loop to achieve multi-fold bandwidth expansion. For system compactness, the feedback loop and RFS loop share a common long fiber. The combined operation produces a broadband swept optical signal, which, through stimulated Brillouin scattering-based frequency-to-time mapping, enables TFA and frequency measurement of broadband microwave signals. Experimental results demonstrate an instantaneous analysis bandwidth of up to 57 GHz, a frequency resolution of 60 MHz, and a maximum mean absolute frequency measurement error of 38.16 MHz.

## 1. Introduction

Time–frequency analysis (TFA) of microwave signals can extract time-varying frequency information from unknown microwave signals and is of great importance in fields such as wireless communications [1] and electronic warfare [2]. To meet the demand for broadband and fast dynamic signal analysis, microwave photonic TFA has attracted extensive attention by taking advantage of microwave photonics, including ultrawide bandwidth, real-time capability, immunity to electromagnetic interference, and reconfigurability [3]. Frequency-to-time mapping (FTTM)-based microwave photonic TFA methods can generally be divided into FTTM via dispersion and FTTM via swept optical signal and filtering. Although the former can achieve nanosecond-scale temporal resolution and analysis bandwidths from tens to hundreds of GHz [4-6], such methods usually require a large amount of dispersion, as well as high-speed, large-bandwidth detection and sampling hardware, resulting in relatively high system complexity and cost.

Compared with FTTM via dispersion, FTTM via swept optical signal and filtering offers significant advantages in frequency resolution and reconfigurability. For example, an all-optical short-time Fourier transform (STFT) scheme based on stimulated Brillouin scattering (SBS) achieved TFA over a 12-GHz analysis bandwidth [7], while an optical chirp chain (OCC) transient SBS scheme realized microwave frequency identification over a 20-GHz measurement range [8]. However, the instantaneous analysis bandwidth of these schemes is still limited by the bandwidth of the swept signal generated by the electrical source. In [9], a frequency shift loop was employed to broaden a 10-GHz-bandwidth linearly frequency-modulated (LFM) signal to 80 GHz. An SBS-based optical filter was subsequently used to construct an FTTM module, enabling instantaneous frequency measurement over a range of 0–40 GHz. However, such schemes typically depend on high-speed, large-bandwidth electrical sweep sources, which increase system cost and hinder miniaturized integration.

To address this issue, the direct generation of high-speed wideband swept optical signals in the optical domain has attracted growing attention owing to its cost-effectiveness. In [10], a frequency-swept optical signal was generated by directly modulating a distributed feedback (DFB) laser with a low-frequency periodic current, thereby enabling STFT, but its analysis bandwidth and measurement performance are limited by the laser modulation response and optical power fluctuations. Period-one (P1) oscillation in an optically injected semiconductor laser can generate tunable frequency-swept optical signals with a high chirp rate and a sweep bandwidth of several GHz or greater [11–13]. TFA based on P1 oscillation has demonstrated signal analysis over a 10-GHz bandwidth [14], and optoelectronic feedback was further used to stabilize the swept signal based on P1 oscillation, producing clearer time–frequency diagrams and reducing the mean absolute error of frequency measurement by about 50% [15]. However, the instantaneous analysis bandwidth is still limited by the sweep bandwidth of the P1 oscillation.

In this letter, we propose and experimentally demonstrate a broadband microwave photonic TFA system based on stabilized P1 oscillation and recirculating frequency shifting (RFS), eliminating the need for expensive wideband electrical sweep sources. A stable and linear swept optical signal is generated from the P1 oscillation of an optically injected semiconductor laser via an optoelectronic feedback loop with iterative predistortion, and its bandwidth is extended by an RFS loop sharing the same long fiber for compactness. The combined operation produces a broadband swept optical signal, which, through SBS-based FTTM, enables TFA and frequency measurement of broadband microwave signals. Experimental results demonstrate an instantaneous analysis bandwidth of up to 57 GHz, a frequency resolution of 60 MHz, and a maximum mean absolute frequency measurement error of 38.16 MHz.

## 2. Principle and Experimental Results

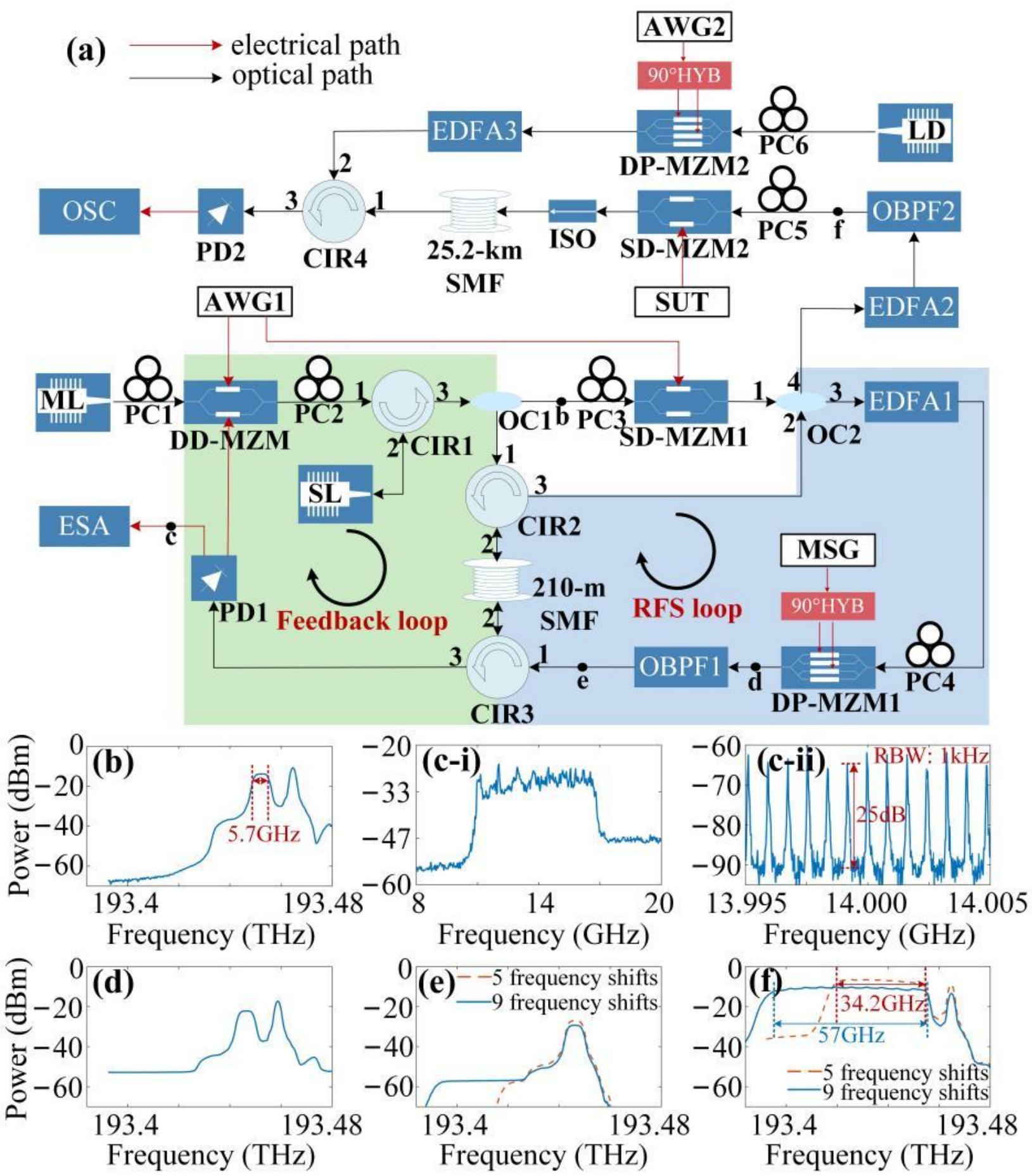

Fig. 1. (a) Schematic diagram of the proposed broadband microwave TFA system. (b) Optical spectrum after optical injection. (c-i) Electrical spectrum of the swept electrical signal output from PD1 under the FDML condition and (c-ii) its enlarged view. (d) Optical spectrum after the first RFS. (e) Spectra at the output of OBPF1 after the first RFS for target RFS numbers of 5 and 9. (f) Spectra of the broadband frequency-swept optical signal after OBPF2 for target RFS numbers of 5 and 9.

Fig. 1(a) shows the schematic diagram of the proposed system. A continuous-wave light emitted from the master laser (ML, ID Photonics CoBriteDX1-1-C-H01-FA) is intensity-modulated at a dual-drive Mach–Zehnder modulator (DD–MZM, Fujitsu FTM7937EZ200), which is driven by a periodic sawtooth control signal generated from arbitrary waveform generator 1 (AWG1, Rigol DG2052). The modulated optical signal is then injected into the slave laser (SL, DFB-1550-10) via a circulator (CIR1). Under appropriate optical injection conditions, the SL is driven into the P1 oscillation state and generates a swept optical signal with a bandwidth of $B_0$ in response to the variation of the injected optical power, whose optical spectrum is shown in Fig. 1(b). Half of the SL output swept optical signal is fed back to the DD-MZM through optical coupler 1 (OC1), CIR2, a section of 210-m single-mode fiber (SMF), CIR3, and a photodetector (PD1, u2t

MPRV1331A). When the period of the sawtooth control signal $T$ matches the feedback loop delay, the Fourier-domain mode-locking (FDML) condition is satisfied, allowing the swept signal to be synchronously fed back and thereby improving its stability. The quality of the swept optical signal is evaluated by monitoring the electrical spectrum of the swept signal output from PD1, which is shown in Fig. 1(c). A clear comb-like electrical spectrum with a comb contrast ratio above 25 dB can be observed when the FDML condition is satisfied. Subsequently, the sawtooth control signal is predistorted using an iterative predistortion method [16] to compensate for the sweep nonlinearity.

The other part of the swept optical signal from OC1 is gated by a single-drive Mach–Zehnder modulator (SD-MZM1, Fujitsu FTM7938EZ) driven by a square wave also generated by AWG1, and synchronized with the sawtooth wave before entering the RFS loop. The square waveform has a period of $(N+1)T$ and a duty cycle of $1/(N+1)$, where $N$ is the target RFS number. Therefore, in every $N+1$ period of the periodic swept optical signal output by SL, only the first period is input into the RFS loop. The gated signal is injected into a 2×2 OC (OC2), and one output is amplified by an erbium-doped fiber amplifier (EDFA1, Amonics AEDFA-PKT-DWDM-15-B-FA) before being injected into a dual-parallel Mach–Zehnder modulator (DP-MZM1, Fujitsu FTM7961EX) that is biased as an optical frequency shifter. A single-tone radio-frequency (RF) signal with a frequency equal to the sweep bandwidth $B_0$ is generated by a microwave signal generator (MSG, HP 83752B) and then applied to DP-MZM1 via a 90° hybrid coupler to shift the swept optical signal. The spectrum after the first frequency shift is shown in Fig. 1(d). In the RFS loop, a tunable optical bandpass filter (OBPF1, EXFO XTM-50) is used to suppress the carrier and the spectral components exceeding the preset number of frequency shifts. The output optical signal from OBPF1 after the first RFS corresponding to target RFS numbers of 5 and 9 are shown by the red and blue curves in Fig. 1(e), respectively. Since the target RFS number is different, the filter bandwidth of OBPF1 is set accordingly to minimize the impact of noise accumulation. The optical signal from the OBPF1 is injected into the 210-m SMF via CIR3 in the direction opposite to that in the aforementioned feedback loop, and further coupled into the other input port of OC2 via CIR2 to close the RFS loop. Since the lengths of the feedback loop and the RFS loop are strictly matched, after $N$ frequency shifts, the original swept optical signal and its frequency-shifted replicas are concatenated in both the time and frequency domains, and a broadband swept optical signal with a bandwidth of approximately $(N+1)B_0$ is obtained at the other output of OC2. After this optical signal is amplified by EDFA2 (Amonics AEDFA-PA-35-B-FA) and filtered by OBPF2, the spectra obtained with target RFS numbers of 5 and 9 are shown by the red and blue curves in Fig. 1(f), respectively. As can be seen, the sweep bandwidth of the concatenated swept optical signal is effectively extended as the target number of frequency shifts increases.

The generated broadband swept optical signal is then injected into SD-MZM2, and the signal under test (SUT) is loaded onto the swept optical signal through carrier-suppressed double-sideband (CS-DSB) modulation. The modulated optical signal is injected into a 25.2-km SMF and interacts with the counter-propagating SBS pump. In the pump branch, the optical carrier generated by a laser diode (LD, ID Photonics CoBriteDX1-1-C-H01-FA) is injected into DP-MZM2 and modulated by a narrowband LFM signal generated by AWG2 (Keysight M8190A) in the carrier suppressed single sideband (CS-SSB) format to broaden the pump bandwidth. The broadened pump wave is then amplified by EDFA3 (Max-Ray EDFA-PA-35-B) and injected into the 25.2-km SMF to excite SBS. By tuning the optical carrier frequency generated by the LD and the pump sweep range, the SBS gain position and equivalent SBS bandwidth can be adjusted [17], respectively. After SBS-based FTTM, the output pulses are detected by PD2 (Nortel PP-10G) and recorded by an oscilloscope (OSC, R&S RTO2032). By extracting the pulse positions period by period and reconstructing them in chronological order, the time–frequency distribution of the SUT can be obtained.

In the experiment, the delays of the two loops were first matched to ensure consistency. The measured loop delay was ultimately 1.2397 μs. Subsequently, the repetition frequency of the sawtooth waveform was adjusted to 806.54 kHz to make the feedback loop satisfy the FDML condition. Its peak-to-peak voltage was set to 2.1 V. The ML was operated at 193.465 THz with an output power of 13.5 dBm. Under a bias current of 11.2 mA and a stabilized temperature of 20.451 °C, the free-running frequency and output power of the SL were 193.451 THz and 3.2 dBm, respectively. Under these circumstances, a frequency-swept optical signal with a period of 1.2397 μs and a bandwidth of 5.7 GHz was generated.

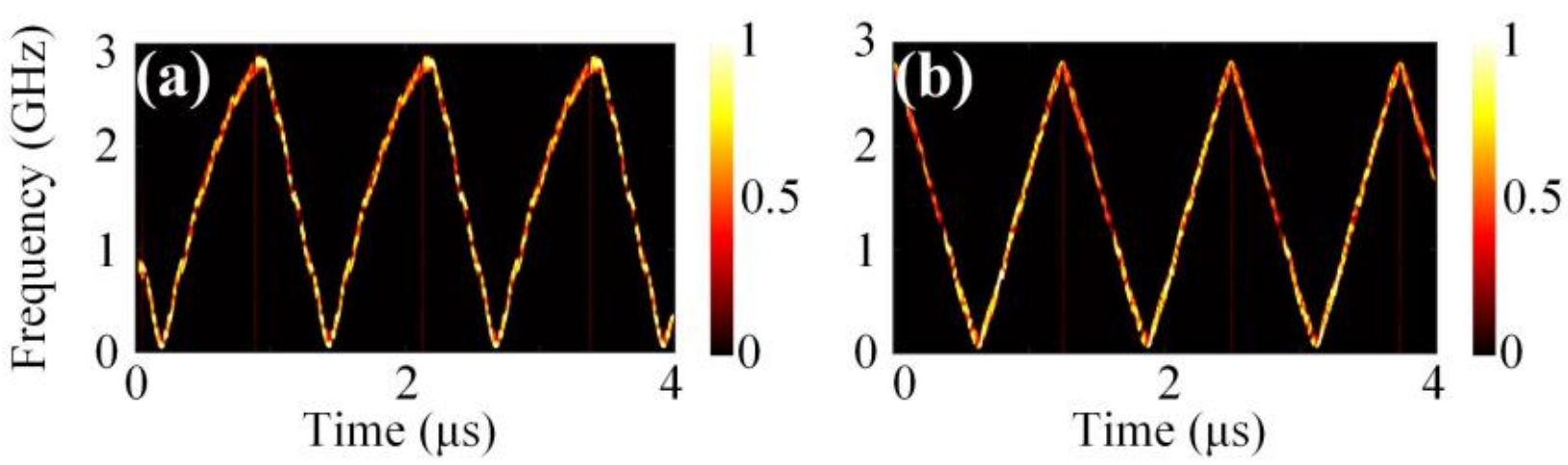

Fig. 2. Time–frequency diagrams of the triangular-chirp electrical signal downconverted from the single-chirp frequency-swept optical signal (a) before and (b) after compensation.

To optimize the sweep linearity, a continuous-wave optical carrier was placed at the center of the swept optical spectrum to down-convert the single-chirp frequency-swept optical signal into a triangular chirped electrical signal, thereby lowering the sampling bandwidth requirement of the OSC. Fig. 2 shows the time–frequency diagrams of the swept signal before and after compensation, respectively. According to [18], the sweep linearity can be defined as the ratio of the maximum frequency deviation to the sweep bandwidth. After four iterations of predistortion, the sweep linearity is improved from $1.27 \times 10^{-1}$ to $7.75 \times 10^{-3}$, which is on the same order of magnitude as the sweep linearity of $4.2 \times 10^{-3}$ reported in [16]. The original sweep bandwidth was obtained from the time–frequency diagram to be approximately 5.7 GHz. Therefore, a 5.7 GHz single-tone RF signal generated by a microwave signal generator is used to drive DP-MZM1 for frequency shifting. For target RFS numbers of 5 and 9, the obtained sweep bandwidths are 34.2 GHz and 57 GHz, respectively. In the pump path, the sweep bandwidth of the pump waveform is set to 70 MHz to achieve an improved measurement performance [17].

To more comprehensively evaluate the TFA capability of the system over a wide bandwidth, the optical carrier frequency output from the LD is adjusted so that the SBS gain spectrum is located at the center of the swept optical spectrum. Meanwhile, the SUT is loaded onto the swept optical signal through CS-DSB modulation, allowing both the positive and negative frequency sidebands to be mapped into temporal pulses through the SBS-based optical filter. This configuration is intended to use the dual-sideband mapping relationship to equivalently cover a larger swept analysis range with lower SUT frequencies, thereby enabling a more complete evaluation of the broadband analysis performance of the system.

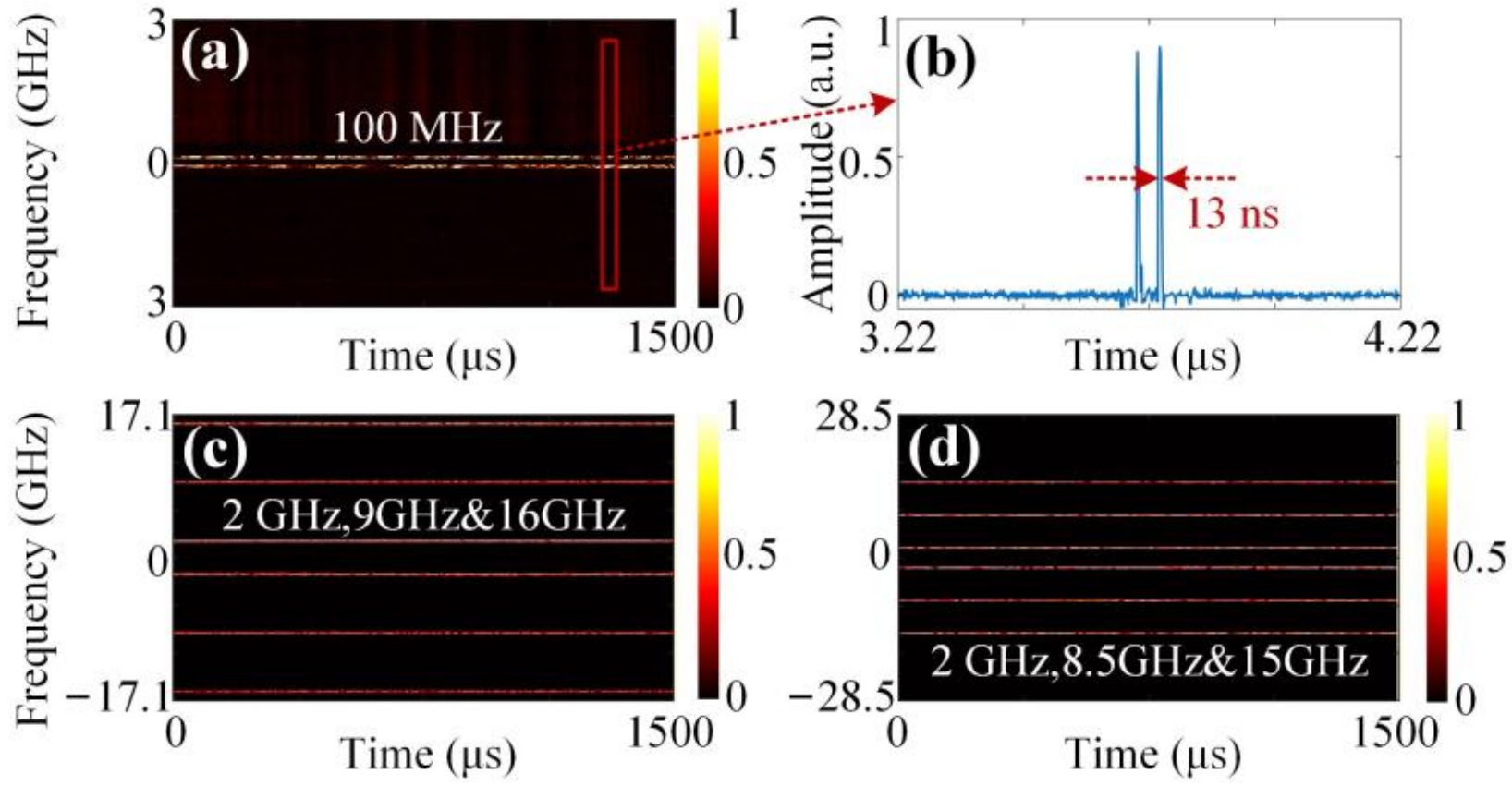

Fig. 3. Measured time–frequency diagrams. (a) 100-MHz single-tone signal with RFS numbers of 5. Three-tone signals at (b) 2, 9, and 16 GHz with RFS numbers of 5 and (c) 2, 8.5, and 15 GHz with RFS numbers of 9.

A 100-MHz single-tone signal was first used as the SUT to evaluate the frequency resolution of the system. As shown in Fig. 3(a), due to CS-DSB modulation, two straight lines appear at 100 MHz and −100 MHz in the time–frequency diagram. Fig. 3(b) shows the pulse waveform within one sweep period. Two pulses associated with the 100 MHz and −100 MHz components are observed. The full width at half maximum of the pulse is around 13 ns. Since the chirp rate is 4.6 GHz/μs, the corresponding frequency resolution is approximately 60 MHz. Subsequently, three-tone signals were employed as the SUT to verify the system's capability in resolving the time–frequency characteristics of signals over a larger instantaneous bandwidth. For a target RFS number of 5, the sweep bandwidth is extended to approximately 34.2 GHz, and the ±2, ±9, and ±16 GHz components corresponding to the 2, 9, and 16 GHz SUT can be observed in Fig. 3(c). For a target RFS number of 9, the sweep bandwidth is further extended to approximately 57 GHz, and the six symmetric frequency components corresponding to the 2, 8.5, and 15 GHz SUT are correctly mapped, as shown in Fig. 3(d). Since one complete broadband sweep is formed over $(N+1)T$, the temporal resolutions are approximately 7.44 μs and 12.40 μs for target RFS numbers of 5 and 9, respectively.

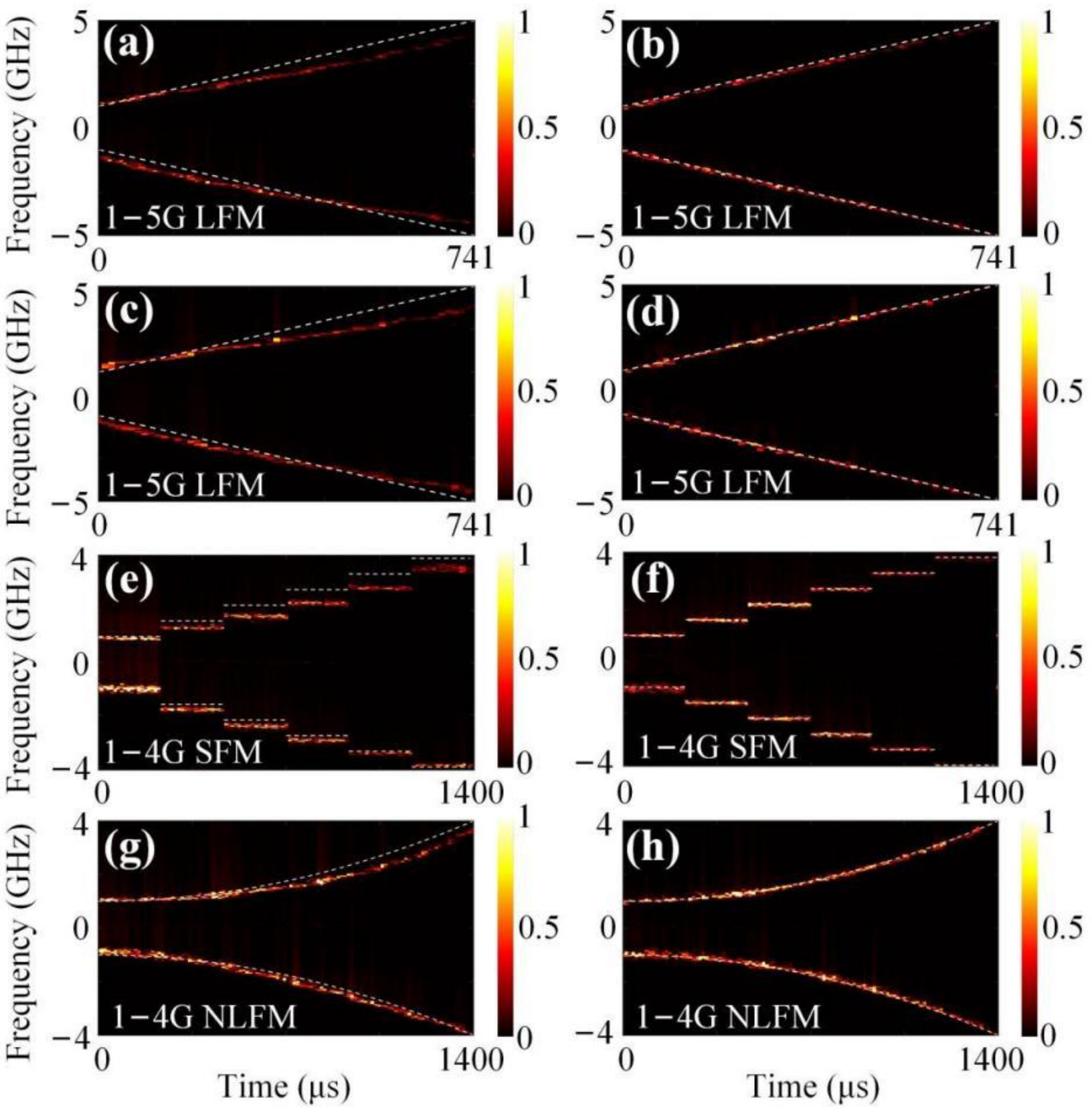

Fig. 4. Measured time–frequency diagrams before and after sweep-linearity compensation. (a), (c), (e), (g) Before compensation; (b), (d), (f), (h) after compensation. The signals are a 1–5 GHz LFM signal with five frequency shifts in (a), (b), a 1–5 GHz LFM signal with nine frequency shifts in (c), (d), a 1–4 GHz step-frequency signal with nine frequency shifts in (e), (f), and a 1–4 GHz NLFM signal with nine frequency shifts in (g), (h).

To investigate the influence of sweep-linearity compensation on the TFA accuracy, an LFM signal with a duration of 741 μs and a frequency range of 1–5 GHz was first measured for target RFS numbers of 5 and 9. The results before and after compensation for a target RFS number of 5 are shown in Fig. 4(a) and 4(b), respectively, while the corresponding results for a target RFS number of 9 are shown in Fig. 4(c) and 4(d), respectively. Before compensation, the reconstructed time–frequency trajectories are bent and deviate from the theoretical dashed lines, indicating that the sweep nonlinearity is directly transferred to the FTTM results. After compensation, the reconstructed trajectories agree well with the theoretical results in white dashed lines, confirming that iterative predistortion can effectively improve the TFA accuracy. To further verify the applicability of the compensation method to different types of microwave signals, a stepped-frequency (SF) signal and a nonlinearly frequency-modulated (NLFM) signal, both with a duration of 1400 μs and a frequency range of 1–4 GHz, are measured for a target RFS number of 9. The TFA results of the SF signal before and after compensation are shown in Fig. 4(e) and 4(f), respectively, while those of the NLFM signal are shown in Fig. 4(g) and 4(h), respectively. After compensation, both the discrete frequency steps and the nonlinear frequency evolution can be well recovered, demonstrating that the proposed system is capable of performing TFA of various complex microwave signals.

Finally, to quantitatively evaluate the measurement accuracy and continuous-measurement stability of the system under bandwidth extension, the proposed system was used to measure the frequency differences of two-tone signals with frequency gaps of 1, 3, 5, 7, 10, and 14 GHz. For each frequency difference, three independent measurements were performed before and after compensation, respectively, with each measurement consisting of 672 consecutive sweep periods. The frequency differences are calculated as the product of the time intervals between the corresponding pulse pairs in the FTTM results and the sweep chirp rate. The measurement errors and corresponding error bars are shown in Fig. 5, where the error bars represent the standard deviations of the measured errors.

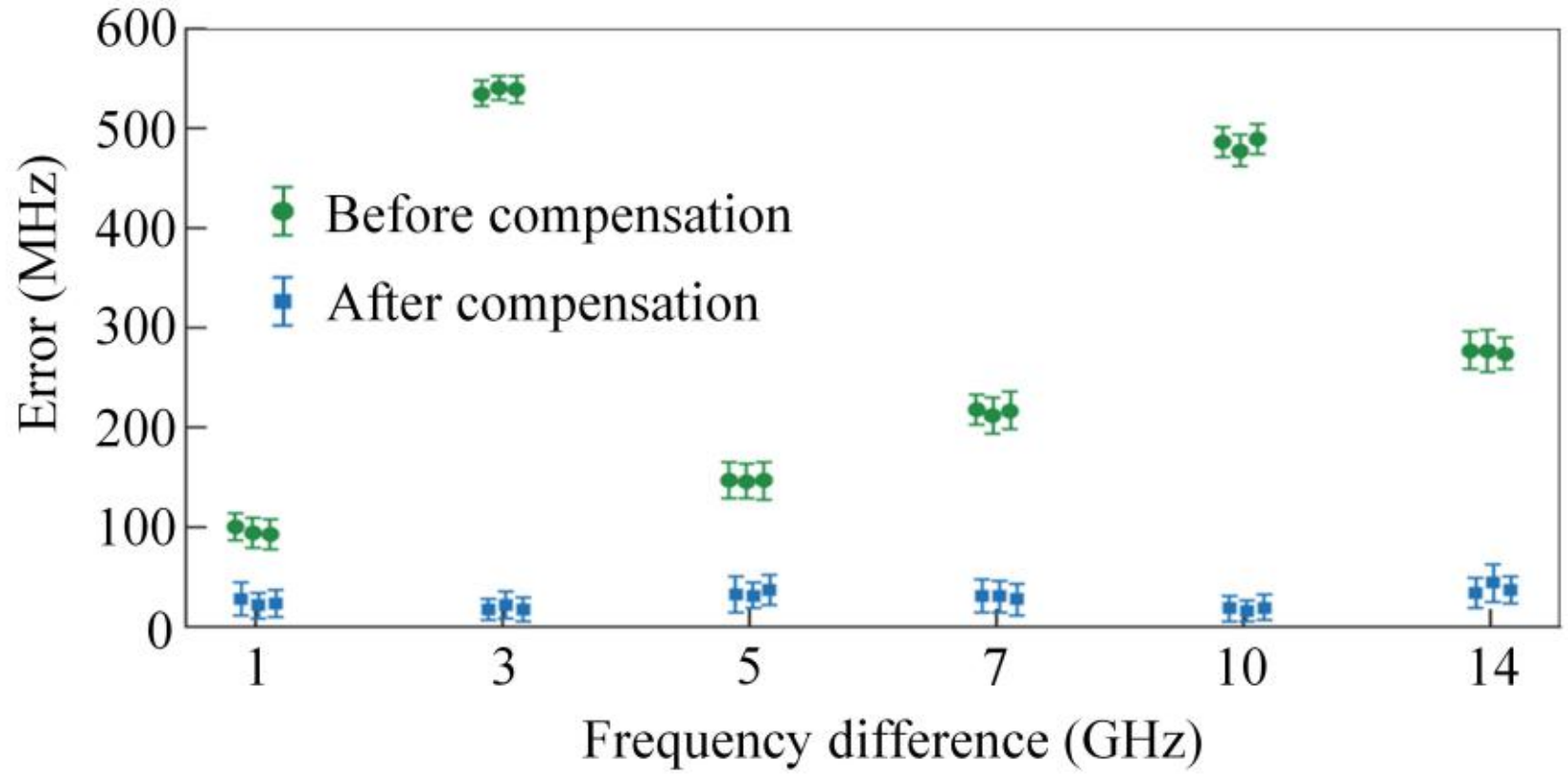

Fig. 5. Statistical frequency-difference measurement errors under a 34.2 GHz analysis bandwidth for 1, 3, 5, 7, 10, and 14 GHz frequency differences before and after sweep-linearity compensation.

Before compensation, the mean absolute errors (MAEs) for different frequency differences are relatively large and vary noticeably with the frequency difference, indicating that sweep nonlinearity directly affects the frequency-difference measurement accuracy. After sweep-linearity compensation, the MAEs for all measured frequency differences are significantly reduced to a very close level. By averaging the three compensated measurements, the average MAEs for the 1, 3, 5, 7, 10, and 14 GHz frequency differences are 24.14, 18.63, 34.02, 29.68, 17.89, and 38.16 MHz, respectively. The relatively small error bars after compensation indicate good stability during consecutive measurements. These results verify that the iterative predistortion method can effectively improve the frequency measurement accuracy of the system when the RFS loop is introduced to broaden the analysis bandwidth.

## 3. Conclusion

In conclusion, we have proposed and experimentally demonstrated a broadband microwave photonic TFA system based on stabilized P1 oscillation and shared RFS and P1-oscillation feedback loop. By combining optoelectronic feedback, iterative predistortion, and RFS, the proposed scheme improves the stability and linearity of the frequency-swept optical signal generated via P1 oscillation while extending its effective sweep bandwidth. By further implementing FTTM via SBS, the system enables broadband TFA of various microwave signals. To the best of our knowledge, this is the first reported work that fuses a P1-oscillation feedback loop with an RFS loop to achieve such a large analysis bandwidth. This work overcomes the problem that the bandwidth of traditional microwave photonic TFA relies on a broadband electronic frequency-swept source, and the problem that the bandwidth of the microwave photonic TFA method based on P1 oscillation is limited by the bandwidth of the optical frequency-swept signal generated by P1 oscillation, providing a promising solution for low-cost broadband TFA.

## Funding

National Natural Science Foundation of China (62371191), Shanghai Oriental Talent Program (QNJY2024007), Shanghai Aerospace Science and Technology Innovation Fund (SAST2022-074), Fundamental Research Funds for the Central Universities, Science and Technology Commission of Shanghai Municipality (22DZ2229004).

## Disclosures

The authors declare no conflicts of interest.

## Data availability

Data underlying the results presented in this paper are not publicly available at this time but may be obtained from the authors upon reasonable request.